\documentstyle[preprint,aps]{revtex}

\newcommand{\lsi}{\,\raisebox{-0.13cm}{$\stackrel{\textstyle<}
{\textstyle\sim}$}\,}
\newcommand{\gsi}{\,\raisebox{-0.13cm}{$\stackrel{\textstyle>}
{\textstyle\sim}$}\,}

\begin{document}
\draft
\preprint{NYU-99-06-02}  
\title{Violation of the Greisen-Zatsepin-Kuzmin cutoff:  
A tempest in a (magnetic) teapot?  \\
Why cosmic ray energies above $10^{20}$ eV may not require new physics
} 
\author{Glennys R. Farrar and Tsvi Piran$^*$
} 
\address{Department of Physics, New York University, NY, NY 10003,
USA\\ 
{$^*$Permanent address:
Racah Institute of Physics, Hebrew University, Jerusalem, Israel}
}

\date{\today}
\maketitle 
\begin{abstract} 

The apparent lack of suitable astrophysical sources for the observed
highest energy cosmic rays (UHECRs) within $\approx 20$ Mpc is the "GZK
Paradox".  We constrain representative models of the extra-galactic
magnetic field structure by Faraday Rotation measurements; limits are at the
$\mu$G level rather than the nG level usually assumed.  In such fields,
even the highest energy cosmic rays experience large deflections.  This
allows nearby AGNs (possibly quiet today) or GRBs to be the source of
ultra-high energy cosmic rays without contradicting the GZK distance limit. 

\end{abstract}

\pacs{
{\tt$\backslash$\string pacs\{\}} }

\narrowtext
%\twocolumn
In 1966, Greisen, Zatsepin and Kuzmin\cite{greisenzatsepin} (GZK)
pointed out that high enough energy protons degrade in energy over
cosmologically small distances due to photopion production from the
cosmic microwave background.  Less than $20\%$ of protons survive
with an energy above $3 \times 10^{20}~(1 \times 10^{20})$ eV for a distance of
18 (60) Mpc\cite{elbert_sommers}.   Ultra-high energy (UHE) nuclei and
photons lose energy even more readily.  Yet more than 10 cosmic rays (CRs)
have been observed with nominal energies at or above $10^{20} \pm 30\%$
eV\cite{AGASA98,watson:HEPiN} with the Fly's Eye event having $3.2
\times 10^{20}$ eV\cite{flyseye}.  Elbert and Sommers, using search
criteria based on the Hillas condition and other reasonable
expectations as to the properties of sources capable of accelerating
protons to these high energies, found no sources within 50 Mpc of
Earth\cite{elbert_sommers}. Thus, it is widely believed that the GZK 
limit must somehow be violated to explain the origin of the observed
UHECRs.  

The need for GZK violation is based on the assumption that UHECRs and
photons from the same source should arrive from the same general
direction and with only a moderate arrival time difference.  If this
assumption is not valid, the observed UHECRs could be produced by
sources within the GZK distance that are not in the direction of the
UHECRs and moreover may have evolved significantly between the UHECR
production and the emission of the photons we now observe.  In this
case, the existance of particles with ultra-high energies would not
raise a paradox if one or more candidates exist for a ``waned" source,
within the GZK distance.  

The coincidence of an observed UHECR and its
astrophysical source depend on the angular deflection of the UHECR
being small.  The UHECR is deflected by the magnetic fields it
encounters which are generally assumed to be of order nG,
leading to expected deflections of order a few degrees or less:
\begin{equation}
\label{angdev}
\delta \theta \sim 0.5^o \sqrt{D_{\rm Mpc} \lambda_{\rm Mpc}} (B_{\rm
nG}/E_{20})  
\end{equation}
being the rms deflection of a proton of energy $E_{20} 10^{20}$ eV
traveling a distance $D$ through randomly-oriented patches of magnetic
field having rms value $B$ and a scale length $\lambda$
\cite{akeno:cluster}.  The corresponding difference between the
arrival times of photons and UHECR's, $\tau_{arr} \sim (\delta
\theta)^2 D/2c$, is small on astrophysical scales, $\approx
10^4$ yr taking $D = L_{GZK} \approx 20$ Mpc for a $E_{20} = 3$ proton
in a nG average field.   With these values both angular and temporal
correlations are expected between the photons and  UHECRs of a given source.  

However, as we show here, extra-galactic magnetic fields (EGMF)'s are
plausibly larger by several orders of magnitude than is generally
assumed, in which case the angular deflection of UHECRs and the time
delay between a UHECR and photons from the same source have been
grossly underestimated.  With larger fields there is no angular and temporal
correlation between a UHECR and its source. The long time delay between
the arrival of UHECRs and of photons implies that the source of the
UHECRS could be unremarkable or even undetectable today.  The expected
cutoff in the UHECR spectrum is modified as discussed below. 

Several groups have recently explored the possibility that the EGMF
has a large scale structure akin to that observed in simulations of
dark matter, assuming mean fields in the 0.1-1 $\mu$G
range\cite{Ryu98,Blasi99}.  Results of our analysis are consistent with
these 
detailed numerical simulations, and clarify the generality of a large
EGMF, $\gsi {\rm few~ tenths}~ \mu$G, relieving the concern that those
results are artifacts of the particular spectra of magnetic
inhomogeneities employed (e.g., Kolmogorov\cite{Ryu98},
log-normal\cite{Blasi99}). 
Subsequent work\cite{Lemoine99,Blasi99a} explored these field ans\"{a}tze
to find what parameters give the best fit to the observed {\it
spectrum} and {\it angular distribution} of CRs above $10^{19}$ eV.  We
instead focus on the puzzling highest energy events, those above
$10^{20}$ eV.  We show that the broad angular distribution and absence
of identifiable sources within the GZK volume is not a problem for
either GRB or AGN sources.  Our discussion focusses on the
interpretation of UHECRs as protons since for nuclei the energy
attenuation distance via photodissociation is shorter than the GZK
length for protons of the same energy.  However our results can be
applied to nuclei as well.   

The structure of this letter is the following. We first determine the
Faraday rotation limits on the EGMFs.  Next we examine the trajectories
of UHE protons in these EGMFs and debunk the notion that the deflection
angles of $10^{20}$ eV protons must be of order a few degrees or less.
This removes the necessity of an angular and temporal correlation
between UHECRs and their astrophysical sources.  Finally we show that
AGNs or GRBs active in the local supercluster in the past 10-100
million years can account for the observed UHECR flux.  We conclude
with a summary and some observational tests of these scenarios.

{\it EXTRAGALACTIC MAGNETIC FIELD:} The prime constraint on the EGMF
arises from the Faraday rotation of light from distant
quasars\cite{Kronberg94}, for which cosmological effects must be 
included.  The rotation measure of a source at a redshift $z$ is 
\begin{equation}
\label{RM}
{\rm RM}(z)=0.2~h_{75}^2 \rm{ \frac {rad}{m^2} }  \int_0^z { n_{05}(z')
B_{||\mu G} (z') \over (1+z')^3} dr_c(z'),
\end{equation}
where $h_{75}$ is the Hubble constant in units of 75km/sec/Mpc,
distances are in Mpc, and $n_{05}$ is the electron density in units of
$(\Omega_b/0.05) (3 H_0^2/8 \pi G)/m_p$;  $\Omega_{b,m,\Lambda}$ are
the ratios of baryonic, matter, and vacuum energy densities to the
closure density. The comoving distance increment, $dr_c$, is related to
the physical distance by $d l = d r_c/(1+z')$.  Additional redshift 
factors arise due to the RM's quadratic dependence on frequency and the
possible cosmological evolution of the electron number density $n_e(z)
\equiv n_{05} (1+z)^{p_e}$ and the magnetic field strength $B(z) \equiv
B_0 (1+z)^{p_B}$.  The cosmological ``dimensionless effective
distance" $d_p(z)$, normalized so that $d_0(z)$ measures the co-moving 
distance in terms of the Horizon distance $ {2 c}/{H_0}$ and with $p
\equiv p_e + p_B -3$, contains the net effect of the cosmological evolution: 
\begin{equation}
d_p(z) \equiv  \int_0^z {dz' (1+z')^p/2 \over {[\Omega_m (1+z')^3 +
(1-\Omega_m - \Omega_\Lambda) (1+z')^2 + \Omega_\Lambda ]^{\frac{1}{2}}}}.
\end{equation} 
In the models below, for $z=2.5$, $\Omega_\Lambda = 1 - \Omega_m$ and $\Omega_m = 1~
(0.3)$, we encounter $d_2 = 1.85 (2.98)$, $d_0 = 0.46 (0.68)$, and
$d_{-3} = 0.14 (0.18)$. 

Faraday rotation measurements for constraining the EGMF make use of
quasars with $\langle z \rangle \approx 2.5$ and yield ${\rm RM} \lsi 5
{\rm ~rad ~ m^{-2} }$\cite{Kronberg94}.  Eqn (\ref{RM}) can be
written ${\rm RM}_5 = 400~ n_{05} ~ B_{||\mu G}~ h_{75}~ d_p(z)$, where
${\rm RM}(z) \equiv 5~{\rm RM}_{5} {\rm~ rad~ m^{-2}}$.  We now
consider two extreme models of the structure of the EGMF.

\noindent {\it Randomly Oriented Patches:}  A commonplace model, which
leads to eqn (\ref{angdev}), posits that the EGMF consists of domains of constant
but randomly oriented field, much like in a ferromagnet, with present
rms strength $B_0$ and characteristic size $\lambda$\cite{Kronberg94}.
Assuming that $n_e$ and $\lambda$ are constant in co-moving
coordinates, and that the energy density of the magnetic field scales
like radiation, implies
\begin{equation}
\label{Brp}
B_0= 0.4 ~ \mu{\rm G} ~ { {\rm RM_5} \over
h_{75}^{3/2} n_{05}} { 1 \over \sqrt{ \lambda_{\rm Mpc}}}{\sqrt{
d_0(z) }\over d_2(z)},  
\end{equation}
where we have taken the mean number of patches in the path to the
quasar to be $r_c(z)/\lambda$.  Assuming $n_e \approx n_b$ and baryonic
closure of the Universe, i.e., $n_{05} \approx 20$, implies nG
fields for $\lambda \approx 1$ Mpc\cite{Kronberg94} and hence the 
small deflections generally assumed in discussions of UHECR
trajectories.  However $n_e$ should be taken at least a factor of 20
smaller than in those estimates, since $\Omega_b \le 
0.05$.  Moreover $\Omega_b$ has contributions from neutrons, and only
electrons in ionized gas are relevant to Faraday rotation, so this
further reduces $n_e$.  Taking the more realistic value $n_{05} \approx
0.3$ gives $B_0$ in this model of order 0.5 (0.4) ${\rm RM_5} (0.3/n_{05})
\mu$G, for the flat $\Omega_m = $ 1 (0.3) cosmologies.

{\it Sheets and Voids:}  It is more likely that the scale structure of
the EGMF consists of randomly oriented sheets of 
field, presumably associated with the sheet-like and filamentary
concentrations observed in the matter distribution, separated by
relative voids.  Idealize this to sheets consisting of layers of
thickness $L_S$ within which the field has a constant magnitude $B_0$
and random orientation,  separated by voids of thickness $L_0 (\approx
50$ Mpc), with $L_S \ll L_0$.  Observations of high redshift clusters
provide strong evidence that $\Omega_m$ is small and the sheets and
voids of matter were largely in place at $z\sim 2.5$\cite{nbahcall}.
Therefore we expect no scaling of $B,~n_e,$ and $L_S$ with redshift in
the sheet-void model, for the redshifts $\le 2.5$ relevant for Faraday
rotation observations.  Thus eqn (\ref{Brp}) can be used to obtain
$B_0$ in this model by replacing $d_2(z)\rightarrow d_{-3}(z)$ and
multiplying by  $\sqrt{L_0/L_S}$.  Note that $B_0$ and $n_e$ stand for
the present rms field strength and electron density {\it within the
sheets} and not averaged over the voids, so $n_e$ can be expected to be
enhanced compared  to the random-patches model by the factor $L_0/L_S$.
Therefore in the sheet-void model we find $B_0$ of order
$d_2(2.5)/(d_{-3}(2.5)\sqrt{50}) $ times that of the random  
patches model.  This is a factor $ \approx 0.9$ in both
cosmologies.  If the sheets themselves consist of many patches of
randomly oriented field of typical size ${\lambda}$, 
$B_0$ should be reduced by a factor $\sqrt{{\lambda}/L_S}$.

The Faraday rotation estimate, $B_0 \approx 0.5~ {\rm RM_5}~ \mu$G,
is only an upper limit.  However, this value is comparable to the
$\approx 1.5\mu$G field argued to exist at the core of the local
(Virgo) Supercluster\cite{Valle90} and to field intensities of $\approx
0.2 ~\mu$G observed in Abell 2319 and Coma (see \cite{Kronberg94} for a
review and \cite{Eilek99}, which appeared after our paper was
submitted, for further evidence for such high magnetic fields.).  A
$\mu$G field corresponds approximately to equipartition between
the magnetic energy and the gravitational and thermal energies of the
supercluster. The consistency of the inferred field from these
disparate approaches increases our confidence in the conclusion: The
EGMF, at least in the local supercluster, is plausibly of order a few
tenths $\mu$G rather than nG as has been traditionally assumed. In the
random patch approximation the energy density of this field is
comparable to that of the CMBR. Therefore we focus primarily on
the more likely ``sheets and voids" model.   

{\it TRAJECTORIES OF UHE PARTICLES:} The Larmor radius of a proton with
an energy $E_{20} 10^{20}$eV in a constant orthogonal field, $B_\perp$, is 
\begin{equation}
\label{RL}
R_L =  0.11~{\rm Mpc} ~ E_{20} / B_{\perp \mu{\rm G}},
\end{equation}
corresponding to a deflection angle of order $ \delta \theta \approx
0.5^o~ B_{\perp \rm nG}~ \lambda_{\rm Mpc}/ E_{20} $ when traversing a
distance $\lambda \ll R_L$.  In the random patches model, we can
express the deflection angle directly in terms of the Faraday rotation
measure allowing the patch-size and its uncertainty to be eliminated.
For a source at a distance $D \gg \lambda$ the relationship is $
\delta \theta_{rms} \approx 60^o~\sqrt{D_{\rm Mpc}}~{\rm RM}_5/(n_{05}~
E_{20}~ h_{75}^{3/2}) $, as long as this deflection 
is small.  An analogous relationship can be obtained for the sheet-void
model, however it is only applicable for cosmic rays which traverse
many sheets, which is not the case for UHECRs since the GZK length is
less than or of order the void size.   

Vall\'ee\cite{Valle90} quotes an average coherent enhancement $\gsi 1.2~
\mu$G in the central 10 Mpc region of Virgo, our local supercluster.
There are surely some inhomogeneities at the 0.1 - 1 Mpc scale, if only
from the galaxies themselves.  In such a field, the minimum energy
above which UHECRs have approximately rectilinear motion over a
distance of 10 Mpc can be estimated by requiring (\ref{angdev}) to
evaluate to $\le  10^o$ with scale size $\lambda = 0.1$ Mpc and $ <B>
= 1.2 \mu$G.  Such small deflection requires $E \gsi 6~10^{21}$ eV,
far larger than in the field configurations considered in refs.
\cite{Lemoine99,Blasi99a}, for which particles above $(1-2)~ 10^{20}$
eV display rectilinear motion.  Given that fields may be larger than
previously assumed, we should not rule out the possibility of magnetic
confinement, diffusive motion, or large angle scattering for even the
highest energy CRs thus far observed.   

{\it IMPLICATIONS FOR UHECR SOURCES:} Now we apply these results to
possible sources of UHECRs. We consider  either continuous
sources (specifically AGNs) or bursting sources (specifically GRBs).
In addition to being capable of accelerating the UHECR to the required
energy the sources must satisfy two additional constraints.
First, the effective number of sources, $N_{eff}$, within the effective
volume contributing to UHECRs on Earth, $V_{eff}$, must be $ \ge 1$. If
$N_{eff} < 1$, it is improbable that a source exists within the GZK
distance during a time interval such that its cosmic rays arrive at
Earth in the past 30 years.  Second, the sources should be energetically
capable of producing  the observed flux of UHECRS on Earth. 

The effective number of active sources within the effective volume
satisfies $ N_{eff} = V_{eff} \times \rho / \epsilon_s$ for constant
sources, and $ N_{eff} = V_{eff} \times \Gamma \tau_{arr}$ for bursting
sources.  Here $\rho$ and $\tau$ are the density and lifetime of active
sources, $\tau_{arr}$ is the spread in UHECR arrival times, $\Gamma$
is the rate per unit volume for bursting sources and $\epsilon_s$ is the
``duty-factor" of a long-lived but not eternal source, e.g., $\epsilon_s
= \tau_{arr}/{\rm min}(\tau,\tau_{arr})$ for an AGN.  The effective
volume, $V_{eff}$, depends on the magnetic field model.  In the
sheet-void model, the contributing region is the local supercluster
within a distance of $L_{GZK}$:  if the cosmic ray experiences a few
large angle scatterings as it travels from source to Earth, $V_{eff}
\rightarrow \pi L_{GZK}^2 D_{LSC}$, where $D_{LSC} \lsi 10$ Mpc is the
thickness of the local supercluster.

The spectrum of UHECRs above $10^{18.5}$ eV can adequately be described
by $E^2 j(E) \approx 3~ {\rm eV} {\rm cm}^{-2} {\rm s}^{-1} {\rm
str}^{-1}$, corresponding to an energy flux per decade of energy,
$\Phi_{obs}$, which is about $3~10^{45} {\rm erg} {\rm Mpc}^{-2} {\rm
yr}^{-1} {\rm str}^{-1}$.  The energy flux per decade in UHECRs,
produced by $N_{eff}$ sources is: 
\begin{equation} 
\label{Eperdec} 
\Phi = \epsilon_{CR} E_{10} \frac{c}{4 \pi} \frac{ \epsilon_s N_{eff}}{
V_{eff}}, 
\end{equation} 
where $E_{10} $ is the energy produced per
energy decade in other forms (e.g. gamma-rays) by the source and
$\epsilon_{CR}$ is the relative efficiency of producing an equal
energy per decade in UHECRs.  For a continuous source, $E_{10} =
P_{10}\tau_{GZK}$, where $P_{10}$ denotes the power per decade and
$\tau_{GZK} = L_{GZK}/c$. 

{\it AGNs and Hot Spots in Giant Radio Galaxies} are the canonical
example of sources that satisfy the Hillas acceleration conditions.
However, there are no suitable AGNs within the GZK distance in the
directions of the observed UHECRS.  If large deflection angles are the
norm, the alignment requirement is relaxed and all powerful radio
sources within the GZK distance become candidates.  Elbert and Sommers 
\cite{elbert_sommers} list eight such high flux radio galaxies.  Among
them M87 (and possibly Cen A) are possibly strong enough to satisfy the
Hillas condition\cite{Biermann87}.  These objects and others in 
the list might also have been more powerful AGNs in the past and thus
are candidate sources\cite{elbert_sommers} in this picture.

The UHECR flux from AGNs is computed using eqn \ref{Eperdec}.  
The power of a moderate AGN, in the energy decade corresponding to gamma
rays, is of order $10^{45} ~{\rm erg~ s^{-1}}$.  $N_{eff}$ in $V_{eff}$
over the past GZK time must be $\ge 1$, as discussed above.  The energy
flux of UHECRs observed on Earth today due to a single AGN in our local
supercluster is therefore $\Phi \approx 10^{48}~{\rm erg ~Mpc^{-2}
~yr^{-1} }~ \epsilon_{CR} \epsilon_s$.  Hence from an energetics
standpoint there is a comfortable margin for
inefficiency or inactivity in the UHECR production by AGNs in the local
supercluster.  

If M87 is indeed powerful enough to accelerate UHECRs
\cite{Biermann87}, then it is possible that it produces all the  
observed UHECRs. The different arrival directions simply reflect
different trajectories in the EGMF.  An alternative scenario is that no
source within the GZK limit (or only M87) is active today, but such sources
have been active in the past.  This implies that the UHECR-producing
lifetime, $\tau_{AGN}$, must be $\lsi \tau_{arr} \sim \tau_{GZK}$.  If
$\tau_{AGN} \gg \tau_{GZK}$, it would be unlikely that a source which  
was active recently enough to have produced an observed UHECR, would no
longer appear active.  On the other hand, if $\tau_{AGN} \ll
\tau_{GZK}$, the only way to have $N_{eff} \ge 1$ is if $\rho_{AGN}
V_{eff} \gg 1$.  If this were the case, we should see many objects
which might have been good sources during the last GZK time, while in
fact we see just a few.  For example, if $\tau_{AGN}/\tau_{GZK} = 0.1$, the 
probability is $\sim 43\%$ that eight sources which have been active
within $\tau_{GZK}$ would all be inactive today, and the probability is
$\sim 38\%$ that 1 out of the eight would be active.  Thus the
likelihood of finding the current situation would be rather large.
Therefore, we conclude that $\tau_{AGN} \lsi \tau_{GZK} \approx 6~
10^7~{\rm yr}$.  Remarkably, independent estimates of AGN lifetimes are
in the $10^{7}-10^{8}$ yr range\cite{tauAGN}, so the AGN source model
survives a highly non-trivial requirement.

{\it GRBs:}  GRBs might be beamed with an opening angle 
$\theta_{GRB} \approx 1/10$.
With such an opening angle we observe only a fraction 
$\theta_{GRB}^{2}/4 $
of all bursts. Let $\Gamma_{GRB}$ be the overall rate of GRBs per 
unit volume.
The observed rate of GRBs,
$ (\theta_{GRB}^{2}/4 )\Gamma_{GRB} \equiv \Gamma_{-9}~10^{-9}~
{\rm yr}^{-1}~ {\rm Mpc}^{-3}$, is independent of beaming and thus fairly
well determined; if gamma ray bursts follow star formation $\Gamma_{-9}
\approx 1$ \cite{Piran99}.  Uniformity requires that the
spread in arrival times of UHECRs from a single source is long compared
to the time between contributing GRB events.  If magnetic fields are of
order nG or less, UHECRs experience small angular deflections.  Only
GRBs beaming toward Earth contribute, $\tau_{arr} \approx 10^4$ yr, and
$N_{eff}$ is marginal\cite{Waxman-Miralda}.

However if the magnetic field is larger and UHECRs experience large
deflections, the beaming fraction suppression factor is removed.  In
addition, $\tau_{arr}$ increases because of the greater path length; it
is replaced by the GZK attenuation time, $\tau_{GZK}$. If the cosmic
ray experiences a few large angle 
scatterings as it travels from source to Earth, $N_{eff} \approx 10^{5}
\Gamma_{-9}$ for $E = 3~10^{20}$ eV, and the number of sources is
adequate to assure our epoch is not unusual.  Eqn \ref{Eperdec} gives
the energy flux per decade in UHECRs.  Taking $E_{10} \approx 10^{52} $
ergs, $L_{GZK} \approx 60$ Mpc (since $E_{20} \approx 1$ dominates the
spectrum of UHECRs) and making the optimistic assumptions that
$(\theta_{GRB}^2/4)^{-1} = 200$ and 
$\epsilon_{CR}$ = 1, gives $\Phi \approx
10^{46} {\rm erg} ~{\rm Mpc}^{-2} {\rm yr}^{-1} {\rm str}^{-1}$.  Thus,
Gamma Ray Bursts are viable, but not by a large factor, from this
standpoint. 

{\it PREDICTIONS OF THIS SCENARIO:}
Several clusters of 2-3 events having energies from $\approx 4 - 30~
10^{19}$ eV have been observed\cite{akeno:cluster}.  If the ultra-high
energy part of the cosmic ray spectrum consists of protons which
experience large deflections due to few-tenth-$\mu$G fields in the local
supercluster, this must be a statistical fluke.  Even if UHECRs are
produced in a handful of locations within the GZK distance, this
initial localization will not survive subsequent deflection, which is
large {\it and strongly energy dependent} without fine-tuning of the
fields.   Thus clustering or the possible directional correlation with
distant compact radio quasars\cite{fb} would be evidence against
the picture advanced here.    

If M87 is the source of all UHECRs, several consequences follow from
its being located near the center of the local supercluster, at a
distance of 20 Mpc which is approximately the GZK distance for a $3
\times 10^{20}$ eV proton.  When better statistics allow the spectrum
to be investigated at higher energy a cutoff in energy should be 
evident in the data. Additionally,  if, as expected,
 the effective scattering distance
is not too small compared with the supercluster dimensions we should 
find an ``in-out" asymmetry with respect to the center of the local 
supercluster.  
Southern Hemisphere detectors, looking
outward with respect to the local supercluster, would be expected to
see a lower UHE flux than observed in the Northern Hemisphere.  Since
not much increase in pathlength can be tolerated for particles
originating near the GZK distance, we could infer in this case that
magnetic fields are not as strong as are allowed by Faraday Rotation
and equipartition limits.  If, instead, GRBs or multiple AGNs are the
source of UHECRs, the energy cutoff would be softer than with a single
source but the in-out asymmetry might still appear if the sources were
concentrated near the supergalactic center.  

A striking aspect of this picture is that observed CRs come mainly from 
the local supercluster:  at ultra-high energy their range is limited by
the GZK effect, while at lower energy they are confined near their
sources by the relatively strong fields.  This may lead to a unified
explanation for the knee and ankle structures in the CR spectrum; a
more complete treatment of the overall spectrum remains to be addressed
in future work.  

In summary, we have shown that the highest energy cosmic rays can
originate in the local supercluster, either from an AGN which 
now may appear past its prime or from gamma ray bursts which have occurred
in the last 10-100 million years.  We presented two distinct models of
the structure of extragalactic magnetic fields, and recalled
observational data on the field in our own and nearby superclusters,
which indicate that extragalatic magnetic fields can be of order
$\mu$G rather than the nG heretofore generally assumed.
This radically changes the picture of ultra-high energy cosmic rays and
implies that even the highest energy cosmic rays observed to date have
strongly bent trajectories such that i) there is no directional
correlation with the source; ii) there is no temporal correlation
between UHECRs and photons from a given source, on time scales relevant
for identifying the sources, even for AGNs, and iii) the power in
UHECRs which can be supplied by the relevant superposition of sources
contributing over a GZK time, $\approx 10^8$ yr, is easily adequate for
AGNs and may barely be adequate for GRBs. 

The research of GRF was supported in part by NSF-PHY-99-96173.  We thank P.
Biermann, C. Heiles, P. Kronberg, and T. Kolatt for valuable contributions.

%\bibliographystyle{unsrt}
%\bibliography{f,cosmo,susy,qcd}

\end{document}